\documentclass[fleqn,usenatbib]{mnras}

\usepackage{mathptmx}

\usepackage[T1]{fontenc}

\usepackage{graphicx}	
\usepackage{amsmath}	
\usepackage{amssymb}	
\usepackage[usenames,dvipsnames]{xcolor} 
\usepackage[normalem]{ulem} 

\newcommand{\eagle}{\mbox{\sc{Eagle}}}

\newcommand{\flares}{\mbox{\sc Flares}}

\newcommand{\spitzer}{\mbox{\it Spitzer}}

\newcommand{\hubble}{\mbox{\it Hubble}}
\newcommand{\jwst}{\mbox{\it JWST}}
\newcommand{\webb}{\mbox{\it Webb}}
\newcommand{\Webb}{\mbox{\it Webb}}

\title[FLARES VI: Colour Evolution]{First Light And Reionisation Epoch Simulations (FLARES) VI: The colour evolution of galaxies $z=5-15$}

\author[Stephen M. Wilkins et al.]{Stephen M. Wilkins$^{1,2}$\thanks{E-mail: s.wilkins@sussex.ac.uk}, 
Aswin P. Vijayan$^{3,4,1}$, 
Christopher C. Lovell$^{5,1}$, 
William J. Roper$^{1}$,\newauthor  
Dimitrios Irodotou$^{6,1}$, 
Joseph Caruana$^{2,7}$, 
Louise T. C. Seeyave$^{1}$, 
Jussi K. Kuusisto$^{1}$, 
Peter A. Thomas$^{1}$\newauthor  
\\
$^{1}$Astronomy Centre, University of Sussex, Falmer, Brighton BN1 9QH, UK\\
$^{2}$Institute of Space Sciences and Astronomy, University of Malta, Msida MSD 2080, Malta \\
$^{3}$Cosmic Dawn Center (DAWN) \\
$^{4}$DTU-Space, Technical University of Denmark, Elektrovej 327, DK-2800 Kgs. Lyngby, Denmark \\
$^{5}$Centre for Astrophysics Research,  School of Physics, Engineering \& Computer Science, University of Hertfordshire, Hatfield AL10 9AB, UK\\
$^{6}$Department of Physics, University of Helsinki, Gustaf Hällströmin katu 2, FI-00014, Helsinki, Finland\\
$^{7}$Department of Physics, University of Malta, Msida MSD 2080, Malta\\
}

\date{Accepted XXX. Received YYY; in original form ZZZ}

\pubyear{2022}

\begin{document}
\label{firstpage}
\pagerange{\pageref{firstpage}--\pageref{lastpage}}
\maketitle

\begin{abstract}
With its exquisite sensitivity, wavelength coverage, and spatial and spectral resolution, the \emph{James Webb Space Telescope} (\jwst) is poised to revolutionise our view of the distant, high-redshift ($z>5$) Universe. While \webb's spectroscopic observations will be transformative for the field, photometric observations play a key role in identifying distant objects and providing more comprehensive samples than accessible to spectroscopy alone. In addition to identifying objects, photometric observations can also be used to infer physical properties and thus be used to constrain galaxy formation models. However, inferred physical properties from broadband photometric observations, particularly in the absence of spectroscopic redshifts, often have large uncertainties. With the development of new tools for forward modelling simulations it is now routinely possible to predict observational quantities, enabling a direct comparison with observations. With this in mind, in this work, we make predictions for the colour evolution of galaxies at $z=5-15$ using the First Light And Reionisation Epoch Simulations (\flares) cosmological hydrodynamical simulation suite. We predict a complex evolution with time, driven predominantly by strong nebular line emission passing through individual bands. These predictions are in good agreement with existing constraints from \hubble\ and \spitzer\ as well as some of the first results from \webb. We also contrast our predictions with other models in the literature: while the general trends are similar we find key differences, particularly in the strength of features associated with strong nebular line emission. This suggests photometric observations alone should provide useful discriminating power between different models and physical states of galaxies.
\end{abstract}




\section{Introduction}\label{sec:intro}

The study of the distant, high-redshift ($z>5$) Universe stands on the cusp of a revolution thanks to the  \emph{James Webb Space Telescope}. \Webb's combination of infrared coverage, sensitivity, and spectroscopic capabilities should ultimately enable the accurate identification of statistical samples of star forming galaxies to $z>10$ \citep{robertson_galaxy_2021}, and the measurement of many key properties including star formation rates, stellar masses, metallicities, and rest-frame optical morphologies.  

A pillar of \Webb's exploration of the distant Universe will be its broadband photometric observations obtained by NIRCam, MIRI, and NIRISS. In cycle 1 alone \Webb\ will acquire $>1\ {\rm deg}^2$ of NIRCam imaging, with the deepest observations approaching 31 mag.  These observations will enable precise measurement of the rest-frame UV luminosity function, particularly at the faint end, allowing the determination of a faint-end turnover. Photometric observations will also enable the measurement of key physical properties such as stellar masses, star formation rates, ages, and dust attenuation. However, the physical properties inferred from photometric observations alone yield large uncertainties \citep[see, e.g.,][]{Whitler2022}, reducing their usefulness in terms of constraining models. One critical cause of this uncertainty is the impact of nebular line emission \citep[see e.g][]{Zackrisson2008, Schaerer2009, Stark2013, Wilkins2013} which can shift predicted colours by up to 1 mag \citet[e.g][]{Wilkins2020}.


However, with the development of sophisticated forward modelling pipelines - which are used to create synthetic observations - \citep[e.g.][]{skirt2015,powderday} it is increasingly possible to directly compare observed and simulated individual galaxies and populations.
While there remain uncertain elements of this modelling - in particular the modelling of dust and nebular emission - we are now at the point where this forward modelling process is comparable, or perhaps even simpler, than the reverse. Efforts to produce synthetic observations are now ubiquitous in the modelling community \citep[e.g][]{Wilkins2013, Trayford2015, Wilkins2016a, vogelsberger_high_2019, FLARES-II} with the ability to make observational predictions across the electromagnetic spectrum incorporating a range of physical processes, including nebular emission from H\textsc{ii} regions \citep[e.g.][]{Wilkins2013, Orsi2014, Wilkins2020, FLARES-II} and the ISM \citep[e.g.][]{Lagache18, Katz2019, Popping2019, Leung2020, Kannan2022}, and dust attenuation \citep[e.g.][]{Bluetides_dust, vogelsberger_high_2019, FLARES-II, FLARES-III} and emission \citep[e.g.][]{Bluetides_dust, Ma2019, Lovell21, FLARES-III}. 

Beyond integrated photometry/spectroscopy it is now also possible to produce synthetic imaging \citep[e.g.][]{Trayford2015, Snyder2015, Ma2018, FLARES-IV, Marshall2022}, enabling self-consistent comparisons of morphological metrics. Forward modelling has been applied to models covering a wide range of scales, from high-resolution simulations of individual halos \citep[e.g.][]{Ma2019} to the construction of synthetic lightcones encompassing hundreds of square arcminutes \citep[e.g.][]{Williams2018, Laigle2019, Davidzon2019, Somerville2021, Yung2022, Drakos2022}, enabling a direct comparison with observed galaxy populations. 

In this work we make predictions for the redshift and luminosity evolution of \hubble, \spitzer, and \jwst\ colours at $z \geqslant 5$ using the First Light And Reionisation Epoch Simulations \citep[\flares;][]{FLARES-I, FLARES-II} cosmological hydrodynamical simulation suite. The current \flares\ simulations combine the $z=0$ validated \eagle\ \citep[][]{schaye_eagle_2015, crain_eagle_2015} physics model with an innovative simulation strategy, resulting in a large effective volume, and thus dynamic range, of stellar masses and luminosities. 

This article is organised as follows: we begin, in Section \ref{sec:theory}, by exploring predictions for the evolution of NIRCam and MIRI colours using a simple toy model. In Section \ref{sec:flares} we then briefly describe the \flares\ project before, in Section \ref{sec:predictions}, presenting predictions including a comparison with existing observations (\S\ref{sec:predictions.observations}) and other models (\S\ref{sec:predictions.othermodels}). Finally, in Section \ref{sec:conc} we present our conclusions.

\section{Theoretical Background}\label{sec:theory} 

To obtain an understanding for the physical effects that drive the colour evolution of galaxies in the distant Universe, in this section we explore predictions from a simple toy model utilising simple star formation and metal enrichment histories and dust modelling.

In this toy model composite spectral energy distributions (SEDs) are created by combining age/metallicity SED grids with a parametric star formation and metal enrichment history. Specifically, we employ the same stellar population synthesis model (SPS): version 2.2.1 of BPASS: Binary Population And Spectral Synthesis \citep[BPASS;][]{BPASS2.2.1}, and initial mass function (IMF): \cite{chabrier_galactic_2003}, as used by \flares. To account for nebular continuum and line emission we process the pure stellar SED grids using the \texttt{cloudy} photoionisation model \citep{Cloudy17.02}. This follows the same approach as \citet{Wilkins2020} and that utilised by \flares. In short, each pure stellar SED is associated with a H\textsc{ii} region with the same metallicity with scale solar composition, a covering fraction of $1$, and assuming a reference\footnote{This defines the ionisation parameter at the reference age (1 Myr) and metallicity ($Z=0.01$). The ionisation parameter at other ages and metallicities are scaled according to the ionising photon luminosity.} ionisation parameter of $\log_{10}U=-2$. A key difference between this modelling and \flares, however, is the treatment of dust. In the absence of spatially resolved stellar populations and dust distributions in the toy model, we assume a simple screen model, while \flares\ employs a line-of-sight model, in principle assigning a unique attenuation to every star particle. 

\begin{figure}
 	\includegraphics[width=\columnwidth]{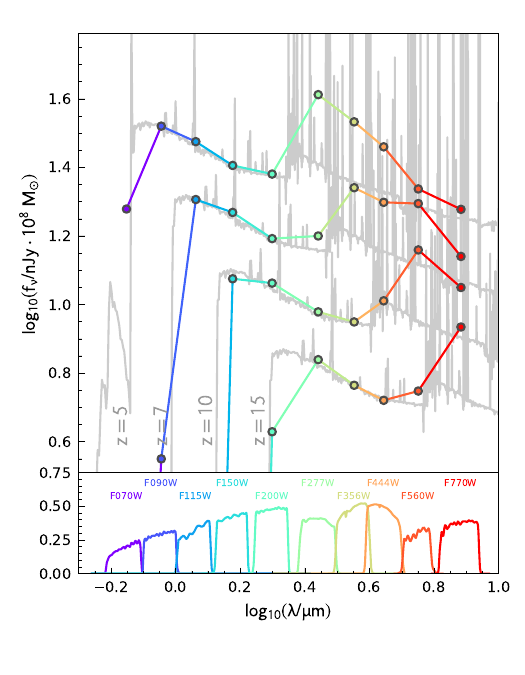}
	\caption{The observed spectral energy distribution of a star forming galaxy at $z=5-15$ alongside key \jwst/NIRCam, and \jwst/MIRI filter transmission functions. Coloured points denote the predicted fluxes in each of the NIRCam and MIRI bands, highlighting the impact of nebular emission in the rest-frame optical. 
	\label{fig:sed}}
\end{figure}

\begin{figure}
 	\includegraphics[width=\columnwidth]{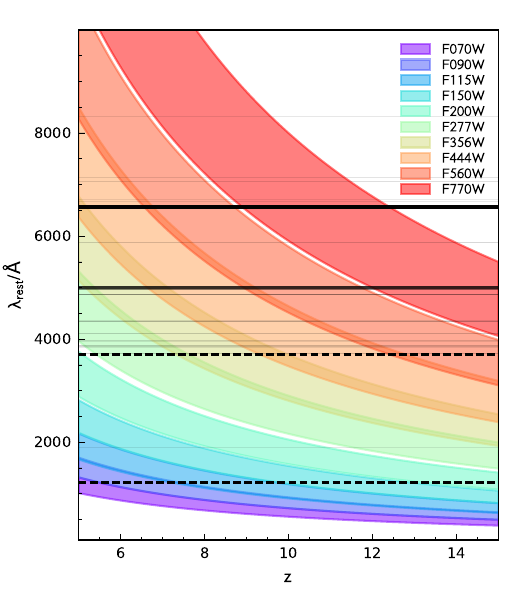}
	\caption{The rest-frame wavelength probed by selected NIRCam and MIRI filters at $z=5-15$. The two dashed horizontal lines denote the location of the Balmer (3646\AA) and Lyman-$\alpha$ (1216\AA) break while solid lines denote strong nebular emission lines with line thickness and opacity indicating the equivalent width for a simple star forming model. 
	\label{fig:rest_wavelength}}
\end{figure}

We begin by using this model to generate spectra of an unobscured (i.e. $\tau_V=0$) composite stellar population, with mass $M_{\star}=10^{8}\ {\rm M_{\odot}}$, a 100 Myr continuous star formation history, and metallicity Z$=0.001$ at $z\in \{5,7,10,15\}$, shown in Figure \ref{fig:sed}. For this fiducial model we assume $f_{\rm esc, LyC}=0$, i.e. maximising the contribution of nebular emission. On this figure we also add predicted broadband fluxes for each of the NIRCam wide filters (F070W, F090W, F115W, F150W, F200W, F277W, F356W, F444W) in addition to the MIRI F560W and F770W bands. Immediately evident in this figure is the impact of the Lyman-limit/$\alpha$ break, the largely smooth UV continuum, and the impact of strong nebular line emission, particularly from [O\textsc{ii}]$\lambda\lambda 3726,3729$\AA,  [O\textsc{iii}]$\lambda 5007$\AA, and H$\beta$. To further aid in the understanding of these predictions, in Fig. \ref{fig:rest_wavelength} we show the rest-frame wavelength probed by the same NIRCam and MIRI filters as a function of redshift, highlighting key emission lines and spectral features. This further reinforces that strong line emission will play an important role in driving NIRCam and MIRI colours at high redshift.

\begin{figure}
 	\includegraphics[width=\columnwidth]{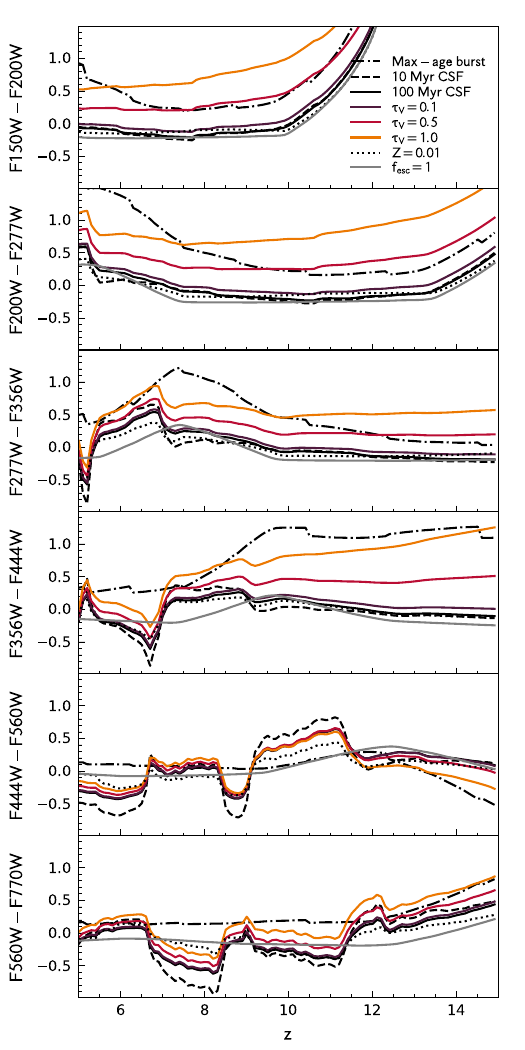}
	\caption{Predicted colours for a range of simple models. Our default model, denoted by the solid black line assumes 100 Myr constant star formation, nebular emissions assuming $f_{\rm esc}=0$ and no dust. Coloured solid lines show the same model but with increasing amount of dust attenuation. The solid grey line instead assumes no nebular emission ($f_{\rm esc}=1$). The dashed line assumes 10 Myr constant star formation. The dotted line assumes $Z=0.01$. The dot dashed line denotes a maximal-aged burst of star formation; effectively the reddest intrinsic colour possible.
	\label{fig:theory_colours}}
\end{figure}

In the following sections we discuss the implications of different model components on the color evolution shown in Figure \ref{fig:theory_colours}.

\subsection{Impact of nebular emission}

Firstly, as denoted by the grey line, we show our fiducial model (100 Myr constant star formation, Z$=0.001$) but with no reprocessing by dust (i.e. $\tau_V=0$) or gas (i.e. $f_{\rm esc, LyC}=1$) - that is, pure stellar emission. The resulting colours evolve smoothly remaining relatively blue ($A-B\approx 0$), except when encompassing the Lyman or Balmer breaks (e.g. F277W-F356W at $z\approx 7.5$) with the latter shifting the colour by up to $\approx 0.5$ mag. We next show our fiducial model (as previous, but with $f_{\rm esc}=0$) as the solid black line. The result is a complex colour evolution with rapid changes, coinciding with strong line emission falling within one of the bands. In this model colours shift by up to $0.5$ mag relative to pure stellar colours. 
This can also lead to colours changing by up-to $0.7$ mag across small redshift intervals (e.g. F444W-F560W at $z\approx 9$). As we will see in Section \ref{sec:predictions} these shifts are even more pronounced when medium and wide filters are combined (e.g. F430M-F444W) with shifts up to $1$ mag predicted for \flares\ galaxies. Where the colour is probing the UV continuum the result is also a shift to redder colours, caused by nebular continuum emission \citep{MassiveBlack_UVslopes}. 

\subsection{Star formation and metal enrichment history}

We next consider models with different star formation histories: a short 10 Myr episode of continuous star formation and a maximal-aged burst\footnote{That is, a model in which the stellar age is equivalent to the age of the Universe at that redshift.}. For the short burst the result is to further enhance the impact of nebular line emission, due to the increased ratio of ionising to optical photons for younger stellar populations. In this model colours now change by up to $1$ mag over short redshift intervals.  
For the maximal-aged burst the resulting colours are consistently redder, due to (increasing) lack of massive hot short lived stars. As the ionising photon luminosity drops rapidly in the first few million years \citep{Wilkins2020} the contribution of nebular emission to these models is extremely limited resulting in little rapid evolution of the colour. In the rest-frame UV the shift from our fiducial model is $\approx 0.5$ mag, growing to $\approx 1$ mag when encompassing the age sensitive Balmer break feature. 

As galaxies in \flares\ span a range of metallicities \citet{FLARES-VII}, we also explore the impact of increasing the metallicity of our fiducial model to $Z=0.01$. Increasing the metallicity both makes the pure stellar SED redder but also reduces the ionising photon luminosity, reducing the overall contribution of nebular emission. Changing the metallicity will also impact individual line ratios leading to a complex impact on colours, especially when one or more strong line is present. However, for the most part the effect of metallicity is fairly subtle, shifting colours by $<0.1$ mag.

\subsection{Reprocessing by dust}

Finally, we consider the impact of dust, applying a screen model assuming a simple $\lambda^{-1}$ attenuation law parameterised using the optical depth in the V-band: $\tau_{V}$. Because this simple model leaves nebular equivalent widths unchanged, where the colours are dominated by nebular emission (e.g. F444W-F560W at $z=5-10$) the result is only a weak shift to redder colours. In the rest-frame UV however, the shift from the fiducial model can be more dramatic, with the colours increasing by $\approx 1$ mag for a model with $\tau_V=1$.

Together, the results in this section highlight the significant impact of many modelling assumptions on the colour evolution of galaxies in the epoch of reionisation, and demonstrates how colours can be used as a key constraint on these assumptions.

\section{The First Light And Reionisation Epoch Simulations}\label{sec:flares}

In this rest of this study, we make use of the core suite of simulations from the first phase of \flares: the First Light And Reionisation Epoch Simulations. The core suite and its initial processing are described in \citet{FLARES-I} and \citet{FLARES-II}, while predictions at the redshift frontier ($z>10$) are presented in \citet{FLARES-V}, and we refer the reader to those articles for a detailed introduction. In short, the core \flares\ suite is a set of 40 spherical re-simulations, $14\ h^{-1}\, {\rm cMpc}$ in radius, of regions selected from a large $(3.2\ {\rm cGpc})^3$ dark matter only simulation. The regions selected to re-simulate span a range of environments: (at $z\approx 4.7$) $\log_{10}(1+\delta_{14}) = [-0.3, 0.3]$\footnote{Where $\delta_{14}$ is the density contrast measured within the re-simulation volume size.} with over-representation of the extremes of the density contrast distribution. As demonstrated in \citet{FLARES-I} this strategy allows us to efficiently simulate a much larger dynamic range in mass (or luminosity) than a traditional periodic box for the same computational resources. We adopt the AGNdT9 variant of the \eagle\ simulation project \cite[][]{schaye_eagle_2015, crain_eagle_2015} with identical resolution and cosmology to the fiducial \eagle\ simulation. This allows us to resolve galaxies with stellar masses $M_{\star} > 10^{8} \; \mathrm{M_{\odot}}$ corresponding to intrinsic rest-frame far-UV 
absolute magnitudes of $M_{\rm UV}\lesssim -18.4$ ($L_{\rm FUV}\gtrsim 10^{28}\ {\rm erg\ s^{-1}\ Hz^{-1}}$) and $m\lesssim 29.1$ at $z=10$ ().

\subsection{Spectral Energy Distribution Modelling}

To produce galaxy observables we process the outputs with a custom pipeline with the approach described in 
\cite{FLARES-II}, broadly following the approach developed by \cite{Wilkins2013, Wilkins2016a, Bluetides_dust, Wilkins2020}, with modifications to the dust treatment. In short, we begin by associating each star particle with a \emph{pure stellar} spectral energy distribution (SED) using v2.2.1 of the Binary Population and Spectral Synthesis \citep[BPASS;][]{BPASS2.2.1} stellar population synthesis model assuming a \cite{chabrier_galactic_2003} IMF according to its age and metallicity. We then associate each star particle with a $\mathrm{HII}$ region giving rise to nebular continuum and line emission following the approach detailed in \cite{Wilkins2020}. We then account for the effect of dust, both in the birth clouds of young stellar populations \cite[with age less than 10 Myr, following][that birth clouds disperse along these timescales]{Charlot_and_Fall} and the wider interstellar medium (ISM). For the latter, we employ a simple line-of-sight attenuation model, similar to that described in \citet{Bluetides_dust}, but using the fitting function for the dust-to-metal ratio presented in \cite{Vijayan2019} (Equation 15 in paper, which parameterises the dust-to-metal ratio as a function of the mass-weighted age of the stellar population and the gas-phase metallicity). For the attenuation due to the birth cloud component, we scale it with the star particle metallicity, thus assuming a constant dust-to-metal ratio. For more details see Section 2.4 in \cite{FLARES-II}.

As demonstrated in Fig. \ref{fig:theory_colours}, broadband colours can evolve rapidly due to the presence of strong emission lines and continuum breaks. However, \flares, like many hydrodynamical simulations, only produces outputs at discrete snapshots, in \flares' case being integer redshifts from $z=15\to 5$. To provide continuous redshift coverage we re-compute observed frame colours from the rest-frame SEDs using perturbed redshifts extending $\pm 0.5$ around the snapshot redshift. For example, galaxies in the $z=10$ snapshot are used to produce predicted colours over the redshift range $z=[9.5, 10.5)$. In practice, we re-sample every galaxy 10 times across the redshift interval, though ensure that no galaxy appears more than once in each $\delta z=0.1$ interval. The downside of this approach is that it assumes no evolution in the physical properties of galaxies between snapshots. However, if there was strong evolution between snapshots this would lead to discontinuities at the boundaries between snapshot redshift ranges. As we will see in Figure \ref{fig:luminosity_Webb}, while discontinuities exist they are small.

\section{Predictions}\label{sec:predictions}

\begin{figure*}
 	\includegraphics[width=\columnwidth]{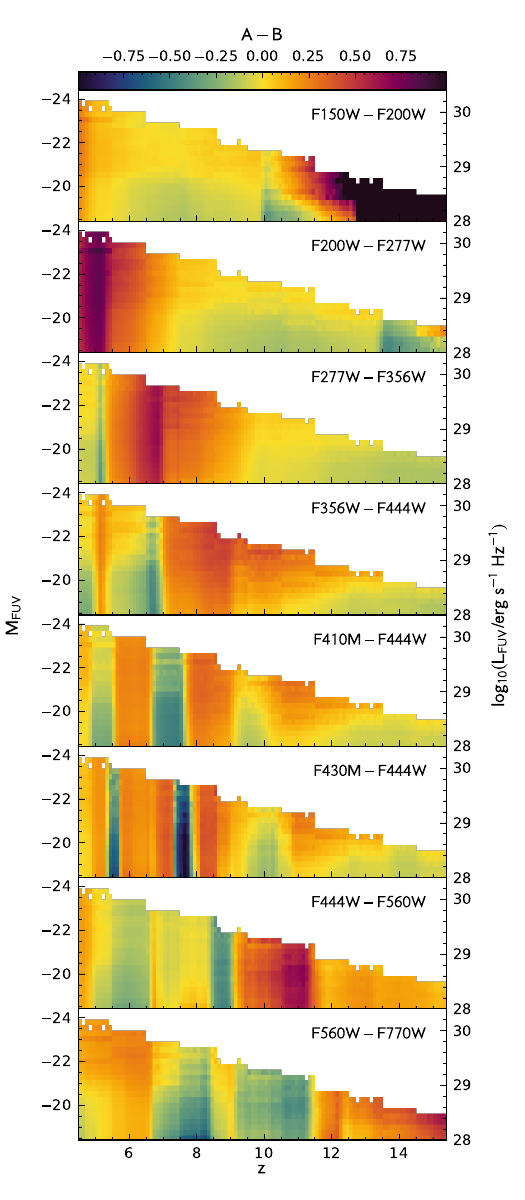}
 	\includegraphics[width=\columnwidth]{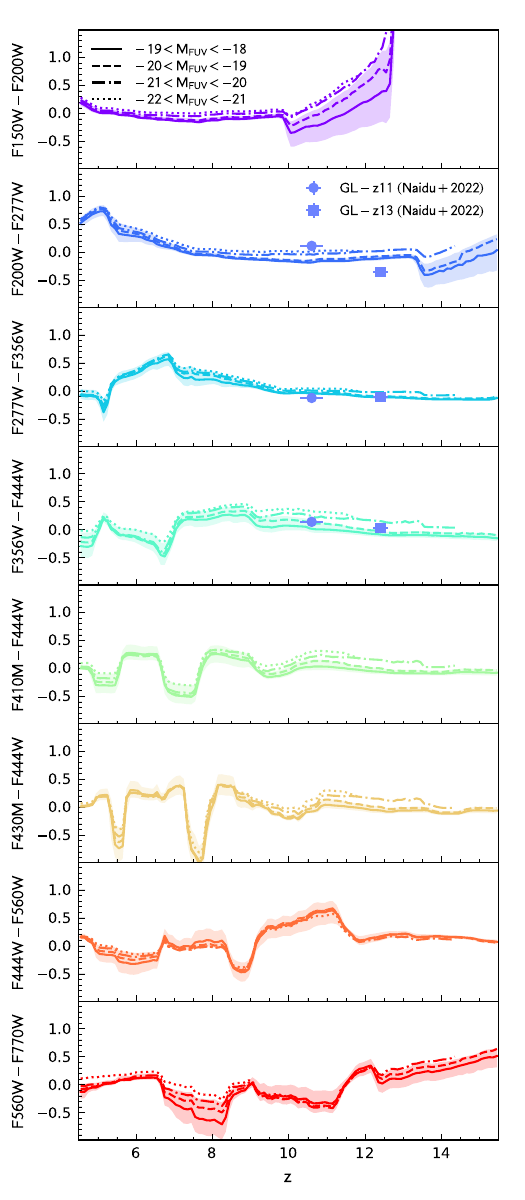}
	\caption{\flares\ predictions for the evolution of NIRCam and MIRI colours of galaxies at $z=5-15$. The left-hand side plot shows the average colour in bins of redshift and rest-frame far-UV luminosity. The right-hand side plot shows both the median colour for galaxies in four absolute magnitude bins and and central 68\% range for the entire sample. Also included in the right-hand side figure are recent observational constraints from \citet{Naidu2022}.
	\label{fig:luminosity_Webb}}
\end{figure*}

We begin by presenting the predictions for the average colour, in various NIRCam and MIRI bands, as a function of redshift and  rest-frame far-UV absolute magnitude in Figure \ref{fig:luminosity_Webb}. Similar predictions for two \hubble/\spitzer\ colours are presented in Figures  \ref{fig:luminosity_Hubble} and \ref{fig:2D_Hubble} in the context of the comparison with current observational constraints described in \S\ref{sec:predictions.observations}. As described in the Data Availability section, we make these results publicly available for comparison with observations or other models.

\begin{figure}
 	\includegraphics[width=\columnwidth]{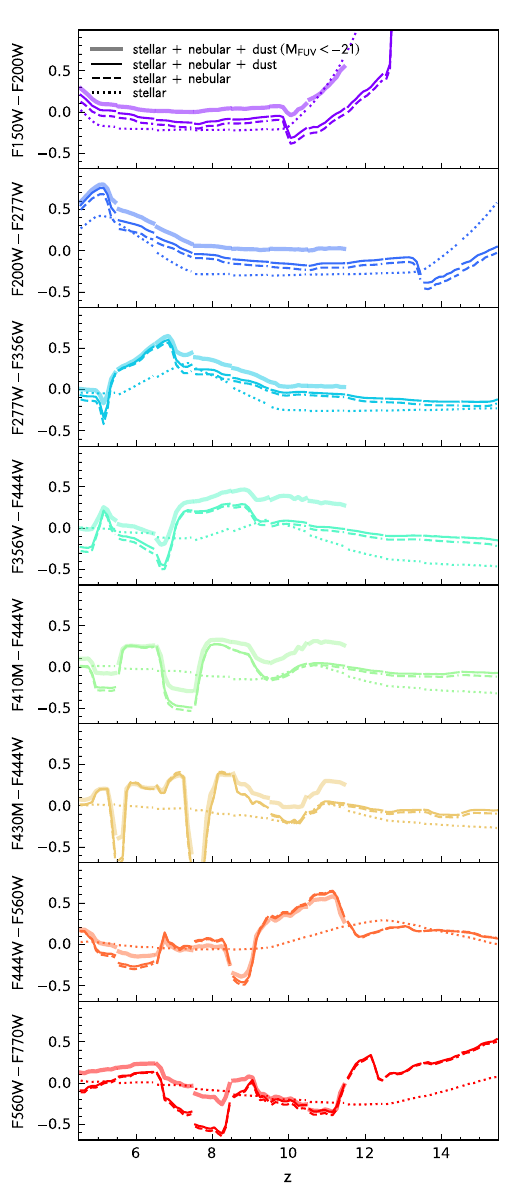}
	\caption{\flares\ predictions for the pure stellar (dotted line), stellar + nebular (dashed line), and observed: stellar + nebular + dust (solid thin line) colour evolution of galaxies with $M_{\rm FUV}<-18$, and in the latter scenario also the brightest ($M_{\rm FUV}<-21$) galaxies (solid thick line).
	\label{fig:type_Webb}}
\end{figure}

Most notable in Figure \ref{fig:luminosity_Webb} is the complex redshift evolution of most colours, with this variability driven by strong nebular line emission and break features (e.g. F150W$-$F200W at $z>10$). To see the impact of nebular line emission more clearly in Figure \ref{fig:type_Webb} we also show the evolution of pure stellar and unattenuated colours. This reveals that the impact of nebular emission is to shift colours by up $0.7$ mag, with the largest shifts for F430M$-$F444W. As strong lines cross to adjacent filters this can also cause sharp redshift evolution of colours, with e.g. F430M-F444W shifting by $>1$ mag from $z=7.5\to 8.5$. This analysis also reveals that nebular emission is often important even in the absence of strong line emission. For example, in colours probing the rest-frame UV continuum (e.g. F356W$-$F444W at $z>12$) the impact of nebular emission reddens colours by up to $0.3$ mag.

Figure \ref{fig:luminosity_Webb} also reveals that the most luminous galaxies are typically redder, albeit the shift is small relative to the variability introduced by nebular emission. As this trend largely disappears for intrinsic colours, we attribute this to the impact of dust attenuation. This is consistent with wider predictions from \flares\ which broadly predicts increasing attenuation with observed UV absolute magnitude \citep[see][]{FLARES-II}\footnote{While there is a broad trend of increasing attenuation with observed UV luminosity, the most heavily attenuated galaxies have more modest UV luminosities due to the effects of attenuation dominating.}

In Figure \ref{fig:luminosity_Webb}, in addition to the median colour, we also show the central 68\% range of colours predicted over the full absolute magnitude range. This range is often very narrow, though does increase to up to 0.7 mag where nebular emission line emission is important. This arises due to the strong sensitivity of the ionising photon luminosity, and thus the contribution of line emission, to the recent star formation histories of galaxies. 

\subsection{Comparison with observations}\label{sec:predictions.observations}

At the time of writing, the first constraints on galaxy colours at high-redshift have just become available \citep{Naidu2022} based on observations from the GLASS and CEERS Early Release Surveys. \citet{Naidu2022} report the discovery of two promising high-redshift galaxy candidates at $z\approx 11$ (GL-z11) and $z\approx 13$ (GL-z11). The reported colours of these sources are included in Figure \ref{fig:luminosity_Webb}. The F277W$-$F356W and F356W$-$F444W colours of both candidates are consistent with the \flares\ predictions. For F200W$-$F277W GL-z11 falls slightly above our predictions and GL-z13 falls slightly below. 

\begin{figure}
 	\includegraphics[width=\columnwidth]{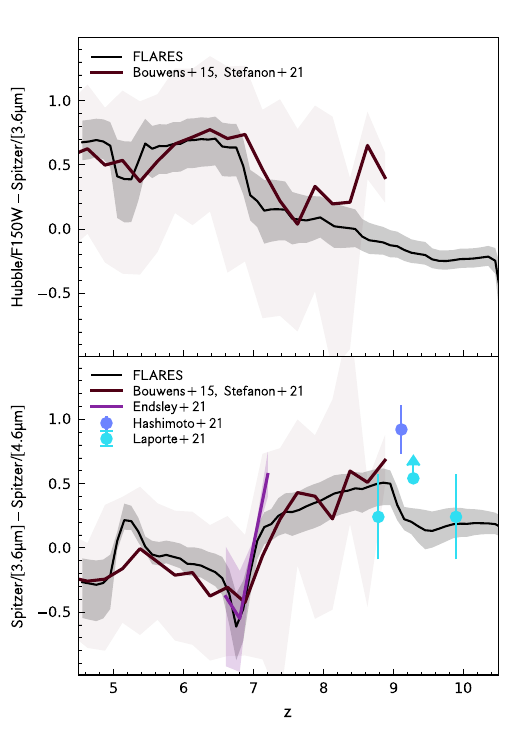}
	\caption{\flares\ predictions, and observations, for the \hubble/WFC3/F160W - \spitzer/IRAC/[3.6] and  \spitzer/IRAC/[3.6] -  \spitzer/IRAC/[4.5] colour evolution of galaxies with $M_{\rm FUV}<-18$. The shaded band shows the central 68\% range of all galaxies $M_{\rm FUV}<-18$. The lines denote the median in the various $M_{\rm FUV}$ bins.
	\label{fig:luminosity_Hubble}}
\end{figure}

The handful of observations available, however, do not provide statistically useful constraints. Instead we compare with comprehensive observed samples \hubble\ and \spitzer. In Figure \ref{fig:luminosity_Hubble} we compare \flares\ predictions for the \hubble/WFC3/F160W - \spitzer/IRAC/[3.6$\mu$m] and  \spitzer/IRAC/[3.6$\mu$m] -  \spitzer/IRAC/[4.5$\mu$m] colours with recent measurements from \citet{Stefanon2021} (based on the sample identified by \citet{Bouwens2015a}), \citet{Endsley2021}, \citet{Hashimoto2018}, and \citet{Laporte2021}. 

\begin{figure*}
 	\includegraphics[width=2\columnwidth]{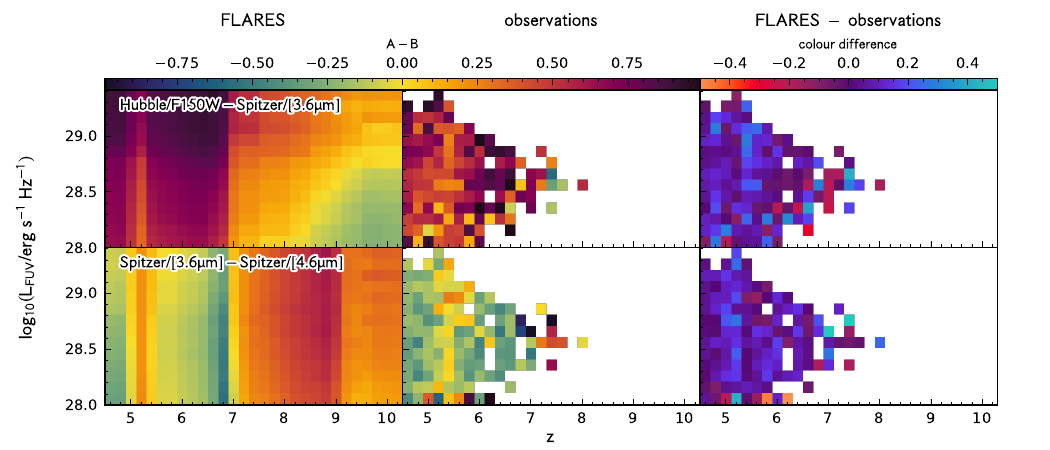}
	\caption{\flares\ predictions for the colour evolution of galaxies with $M_{\rm FUV}<-18$. The shaded band shows the central 68\% range of all galaxies $M_{\rm FUV}<-18$. The lines denote the median in the various $M_{\rm FUV}$ bins.
	\label{fig:2D_Hubble}}
\end{figure*}

\citet{Stefanon2021} consistently reduced \spitzer/IRAC imaging across the Great Observatories Origins Deep Survey (GOODS)-N and GOODS-S fields. This data-set was then used to measure the IRAC fluxes of almost 10,000 galaxies at $3.5<z<10$ based on the catalogue of \citet{Bouwens2015a}. The \citet{Bouwens2015a} sample was selected and assigned photometric redshifts using \hubble\ imaging. The resulting colour evolution of this sample, matched to have the same rest-frame UV luminosity limit ($M_{\rm UV}<-18.4$) is shown in Figure \ref{fig:luminosity_Hubble}. This reveals a median observed colour providing a close match to the \flares\ predictions. The impact of [O\textsc{iii}] and H$\beta$ exiting the IRAC/[3.6$\mu$m] band and entering the [4.5$\mu$m] band at $z=6.5-7.5$ can be clearly discerned. While the median colour is well matched, the scatter in the observations is up to 5$\times$ larger. While photometric scatter and redshift uncertainties will drive some of this difference it is possible this may reflect a real difference between the observations and \flares. While we have applied a consistent luminosity limit to both samples the \citet{Bouwens2015a}/\citet{Stefanon2021} sample do not cover the same range of luminosities. To provide a fairer comparison in Figure \ref{fig:2D_Hubble} we calculate the average colour in bins of luminosity and redshift and contrast \flares\ and the observations. This is inevitably more noisy but doesn't appear to reveal any systematic bias.

In Figure \ref{fig:luminosity_Hubble} we also compare against the sample of \citet{Endsley2021}. \citet{Endsley2021} select a sample of galaxies at $z\sim 6.5-7$ from ground based imaging of the COSMOS and XMM1 fields. They use a colour selection employing Subaru/Hyper Suprime-Cam NB921 narrow-band imaging to yield precise photometric redshifts. This in turn results in clean constraints on the [O\textsc{iii}]+H$\beta$ equivalent widths of the sources from the \spitzer/IRAC photometry. The resulting [3.6$\mu$m] - [4.5$\mu$m] colours at $z\sim 7$ closely match the \flares\ predictions, in particular the very blue colours predicted at $z\approx 6.8$ and strong subsequent evolution to $z>7$. 

\subsection{Comparison with other models}\label{sec:predictions.othermodels}

\begin{figure}
 	\includegraphics[width=\columnwidth]{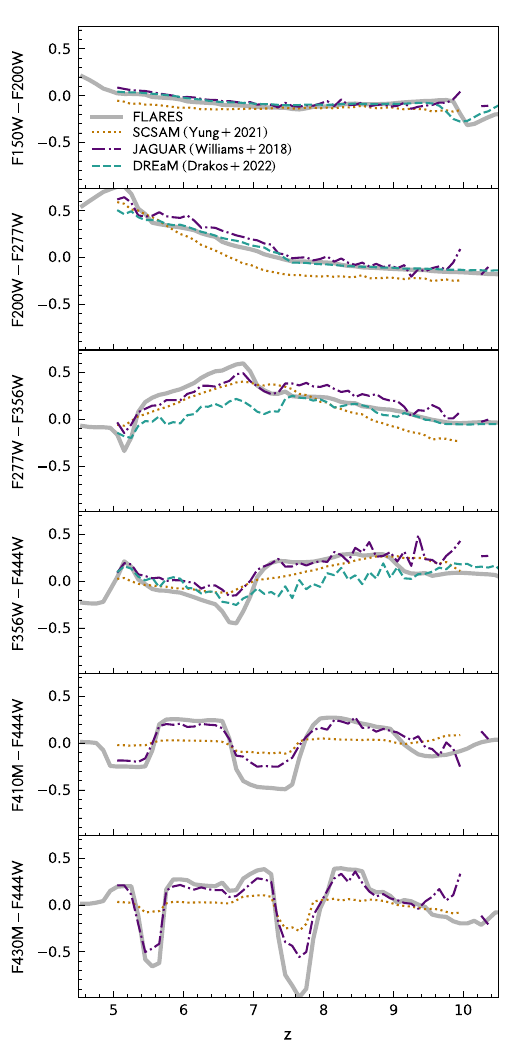}
	\caption{Predictions for the \jwst\ colour evolution of galaxies with $M_{\rm FUV}<-18$ from \flares\ (thick grey line), the Santa Cruz SAM \citep{Somerville2021, Yung2022}, JAGUAR \citep{Williams2018}, and DREaM \citet{Drakos2022}.
	\label{fig:othermodels_Webb}}
\end{figure}

\begin{figure}
 	\includegraphics[width=\columnwidth]{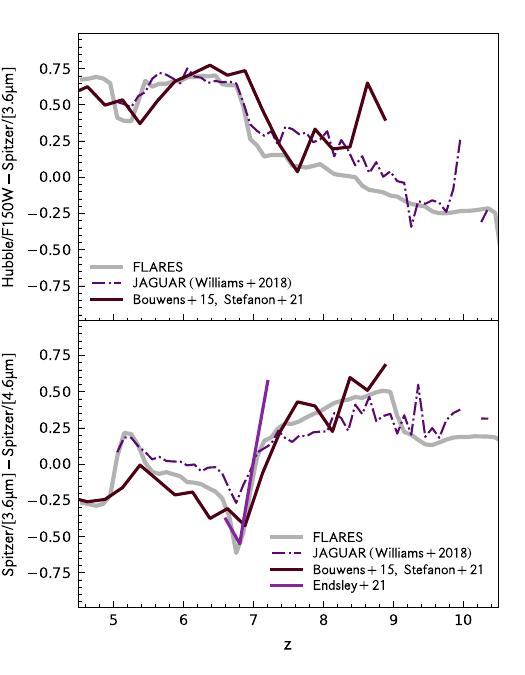}
	\caption{Predictions for the \hubble\ colour evolution of galaxies with $M_{\rm FUV}<-18$ from \flares\ (thick grey line) and JAGUAR \citep{Williams2018}. DREaM and the Santa Cruz SAM are omitted as they do not provide both \hubble\ and \spitzer\ photometry.
	\label{fig:othermodels_Hubble}}
\end{figure}

We now compare our predictions to mock catalogues from the phenomenological models of \citet{Williams2018} (also known as the JAGUAR: JAdes extraGalactic Ultradeep Artificial Realizations package) and DREaM: the Deep Realistic Extragalactic Model \citep{Drakos2022}, in addition to the Santa Cruz semi-analytical model \citep{Somerville2021, Yung2022}. The colour evolution of these models are contrasted with \flares\ in Figure \ref{fig:othermodels_Hubble} and Figure \ref{fig:othermodels_Webb} for \jwst\ and \hubble\ + \spitzer\ colours respectively. 

While the evolution of colours in all three sets of predictions is qualitatively similar there are some important differences, particularly around the impact of nebular emission. Specifically, \flares\ consistently predicts a stronger contribution from nebular emission, resulting in more extreme variation of colours. While, at present, it is not possible to definitively claim one model provides better agreement with the observations, \flares\ does appear to better reproduce the magnitude of the observed dip in the average [3.6$\mu$m] - [4.5$\mu$m] colour at $z\approx 6.8$.

Due to the multitude of differences between the three models, the exact cause of this discrepancy is difficult to ascertain. Possibilities include the presence of more stochastic, young, or rapidly increasing star formation in \flares\ or simply the choice of stellar population synthesis model, initial mass function, and/or photoionisation modelling assumptions \citep[see e.g.][]{Wilkins2013, Wilkins2016, Wilkins2020}.  Critical to ascertaining the cause of this discrepancy is applying a consistent approach to the choice of SPS model, IMF, and photoionisation assumptions should ultimately help diagnose the cause of this discrepancy. At the same time, results from \webb\ are now beginning to provide the observational constraints to differentiate between different sets of predictions. 


\section{Conclusions}\label{sec:conc}

In this work we have presented theoretical predictions for the colour evolution of galaxies at $z=5-15$ from the \flares: First Light And Reionisation Epoch Simulations. These predictions enable direct comparison with observational constraints, allowing both \flares\ and the underlying model to be tested in a new regime. Our major findings are:

\begin{itemize}

    \item The predicted galaxy colours show complex evolution with rapid changes (up to $\approx 1$ mag) in the average colours over short redshift intervals. This is due to the presence of strong nebular line emission moving through individual bands. In addition, our predicted colours show a trend, albeit modest, with luminosity, attributed to the increasing impact of dust in the most luminous galaxies.
    
    \item \flares\ predictions currently closely match, with the possible exception of the scatter, recent observational constraints at high-redshift using \hubble\ and \spitzer\ from \citet{Stefanon2021} \cite[based on the galaxy sample identified by][]{Bouwens2015a} and \citet{Endsley2021}, in addition to early results from \webb\ \citep{Naidu2022}.
    
    \item \flares\ predictions qualitatively match other available predictions including the phenomenological models of \citet{Williams2018} and \citet{Drakos2022} and the \citet{Somerville2021, Yung2022} semi-analytical model. However, \flares\ predicts larger variations due to stronger nebular line emission. The exact cause of this difference may lie in the fundamental physical properties of galaxies in the models but may also reflect different modelling assumptions - e.g. choice of stellar population synthesis model and initial mass function - for the forward modelling.
    
\end{itemize}

With the imminent explosion of constraints from \Webb\ we will soon be in a position to differentiate between these and other models, providing insights into the physics driving the physical and observational properties of galaxies in the early Universe. Mock observations like those presented here, and in e.g. \citet{Williams2018}, \citet{Somerville2021}, \citet{Drakos2022} and \citet{Yung2022}, provide a critical resource to test and refine photometric redshift selection techniques and/or the inference of physical properties from broadband photometry.

\section*{Acknowledgements}

We dedicate this article to healthcare and other essential workers,  the teams involved in developing the vaccines, and to all the parents who found themselves having to home-school children while holding down full-time jobs. We thank Rychard Bouwens, Ryan Endsley, and Mauro Stefanon for providing machine readable catalogues of their observations for comparison. We thank the \eagle\, team for their efforts in developing the \eagle\, simulation code.  This work used the DiRAC@Durham facility managed by the Institute for Computational Cosmology on behalf of the STFC DiRAC HPC Facility (\url{www.dirac.ac.uk}). The equipment was funded by BEIS capital funding via STFC capital grants ST/K00042X/1, ST/P002293/1, ST/R002371/1 and ST/S002502/1, Durham University and STFC operations grant ST/R000832/1. DiRAC is part of the National e-Infrastructure. CCL acknowledges support from the Royal Society under grant RGF/EA/181016. DI acknowledges support by the European Research Council via ERC Consolidator Grant KETJU (no. 818930). The Cosmic Dawn Center (DAWN) is funded by the Danish National Research Foundation under grant No. 140. We also wish to acknowledge the following open source software packages used in the analysis: \textsc{Numpy} \citep{Harris2020_numpy}, \textsc{Scipy} \citep{2020SciPy-NMeth}, and \textsc{Matplotlib} \citep{Hunter:2007}. This research made use of \textsc{Astropy} \url{http://www.astropy.org} a community-developed core Python package for Astronomy \citep{astropy:2013, astropy:2018}. Parts of the results in this work make use of the colormaps in the \textsc{CMasher} package \citep{CMasher}.

\section*{Data Availability Statement}

The 2.2, 15.8, 50, 84.2, and 97.8 percentiles of the colour distribution in redshift and rest-frame far-UV luminosity bins are available in the \texttt{astropy} \href{https://github.com/astropy/astropy-APEs/blob/main/APE6.rst}{Enhanced Character Separated Values (\texttt{.ecsv}) table format} at \url{https://github.com/stephenmwilkins/flares_colours_data} and as part of the wider \href{}{First Light and Assembly of Galaxies} model predictions repository available at \url{https://github.com/stephenmwilkins/flags_data}. Data from the wider \flares\ project is available at \url{https://flaresimulations.github.io/data.html}. If you use data from this paper please also cite \citet{FLARES-I} and \citet{FLARES-II}.



\bibliographystyle{mnras}
\bibliography{flares-colours} 





\bsp	
\label{lastpage}
\end{document}